\documentclass[prl,superscriptaddress,reprint,showpacs]{revtex4-1}

\usepackage{graphicx,textcomp,amssymb,amsmath,dcolumn,hyperref}
\begin{document}

\title{Excited states in bilayer graphene quantum dots}

\author{A.~Kurzmann}
\email{annikak@phys.ethz.ch}
\author{M.~Eich}
\author{H.~Overweg}
\author{M.~Mangold}
\author{F.~Herman}
\author{P.~Rickhaus}
\author{R.~Pisoni}
\author{Y.~Lee}
\author{R.~Garreis}
\author{C.~Tong}

\affiliation{Solid State Physics Laboratory, ETH Zurich, CH-8093 Zurich, Switzerland}
\author{K.~Watanabe}
\author{T.~Taniguchi}
\affiliation{National Institute for Material Science, 1-1 Namiki, Tsukuba 305-0044, Japan}
\author{K.~Ensslin}
\author{T.~Ihn}
\affiliation{Solid State Physics Laboratory, ETH Zurich, CH-8093 Zurich, Switzerland}

\date{\today}

\begin{abstract}
We report on ground- and excited state transport through an electrostatically defined few-hole quantum dot in bilayer graphene in both parallel and perpendicular applied magnetic fields. A remarkably clear level scheme for the two-particle spectra is found by analyzing finite bias spectroscopy data within a two-particle model including spin and valley degrees of freedom. We identify the two-hole ground-state to be a spin-triplet and valley-singlet state. This spin alignment can be seen as Hund's rule for a valley-degenerate system,  which is fundamentally different to quantum dots in carbon nano tubes and GaAs-based quantum dots. The spin-singlet excited states are found to be valley-triplet states by tilting the magnetic field with respect to the sample plane.  We quantify the exchange energy to be $0.35\,\text{meV}$ and measure a valley and spin g-factor of 36 and 2, respectively.
		
\end{abstract}
	
\maketitle
	
\section{Introduction}
Currently, a great variety of physical systems, including trapped ions \cite{cirac1995quantum,kielpinski2002architecture}, superconducting transmons \cite{riste2013deterministic,peterer2015coherence}, and semiconducting quantum dots \cite{petta2005coherent,awschalom2013quantum,loss1998quantum} are competitive implementations of qubits for future quantum information technologies \cite{bennett2000quantum}. Among the materials suitable for high quality semiconductor quantum dots, gallium arsenide \cite{tarucha1996shell,ciorga2000addition,petta2005coherent}, silicon/germanium \cite{hu2007ge,yoneda2018quantum}, and silicon \cite{yang2013spin, lim2011spin,watson2018programmable} are prime candidates, the latter having the great advantage of being compatible with present day processing technologies of semiconductor industry \cite{veldhorst2014addressable}. The key factors limiting qubit coherence in these materials have been identified to be the hyperfine-interaction, spin-orbit interaction, and impurity-related charge-noise \cite{johnson2005triplet,khaetskii2002electron,yoneda2018quantum}. Silicon is believed to minimize most of these detrimental effects, because it can be isotopically purified \cite{veldhorst2014addressable} (minimizing hyperfine interactions), it is a light element (minimizing spin-orbit effects), and is one of the purest technological materials available (minimizing impurity-related charge noise). Experimental evidence is currently growing, that another material system, namely graphene combines similar virtues \cite{trauzettel2007spin,silvestrov2007quantum,wang2007z,sols2007coulomb} and is now reaching the quality to become competitive with silicon \cite{eich2018spin}.
Recent improvements in fabrication technologies for graphene nanostructures, namely, the encapsulation between boron nitride \cite{dean2010boron}, edge-contacting \cite{wang2013one}, graphite back-gates \cite{zibrov2017tunable}, and the use of electrostatic gating of bilayer graphene \cite{oostinga2008gate,allen2012gate}, have leveraged the quality of quantum point contacts \cite{overweg2017electrostatically,B,kraft2018valley} and quantum dots\cite{B,eich2018spin} to such an extent, that few-electron or -hole quantum dots have been realized that are comparable to the best devices in gallium arsenide.

In this paper we aim at establishing the basis for future qubit implementations in graphene quantum dots by carefully studying and identifying the single-particle and many-body ground- and excited states of quantum dots trapping only one or two charge carriers. While the properties of the material bear similarities to carbon nanotubes \cite{jarillo2004electron,huertas2006spin,kuemmeth2008coupling} and silicon \cite{veldhorst2014addressable,friesen2003practical}, because of the two-fold valley and spin-degeneracies, the results of our experiments allow us to propose a remarkably clear level scheme for two-particle spectra, in which the spin- and valley-entanglement, as well as exchange interactions play a crucial role. With this level scheme at hand, future experiments can investigate spin- and valley-coherence and relaxation times \cite{elzerman2004single,veldhorst2014addressable}, which are key parameters to be compared to other material systems.

\begin{figure}
	\includegraphics[scale=1]{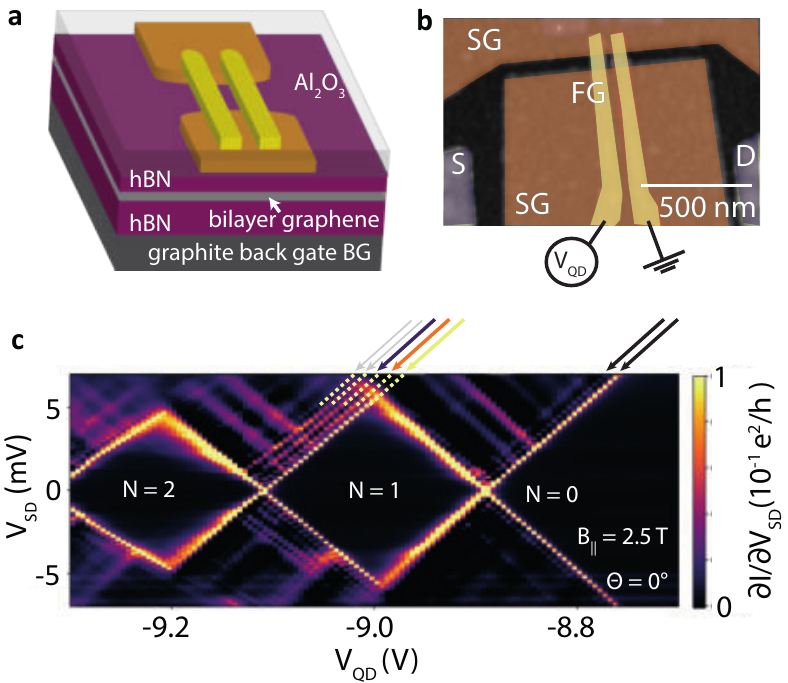}
	\caption{(a) Schematic of the stacked sample with graphite back gate, hBN, bilayer graphene and hBN. (b) False-color atomic force microscope (AFM) image of the device in bilayer graphene. By using the split gates (orange, SG) a conducting channel (black) is created. The finger gates (yellow, FG) across the channel produce quantum dots in the bilayer graphene. (c) Finite bias measurement of the first two Coulomb resonances of a quantum dot in the hole regime for an in-plane magnetic field of 2.5 T.}
	\label{fig1}
\end{figure}

\begin{figure*}
	\includegraphics[scale=1]{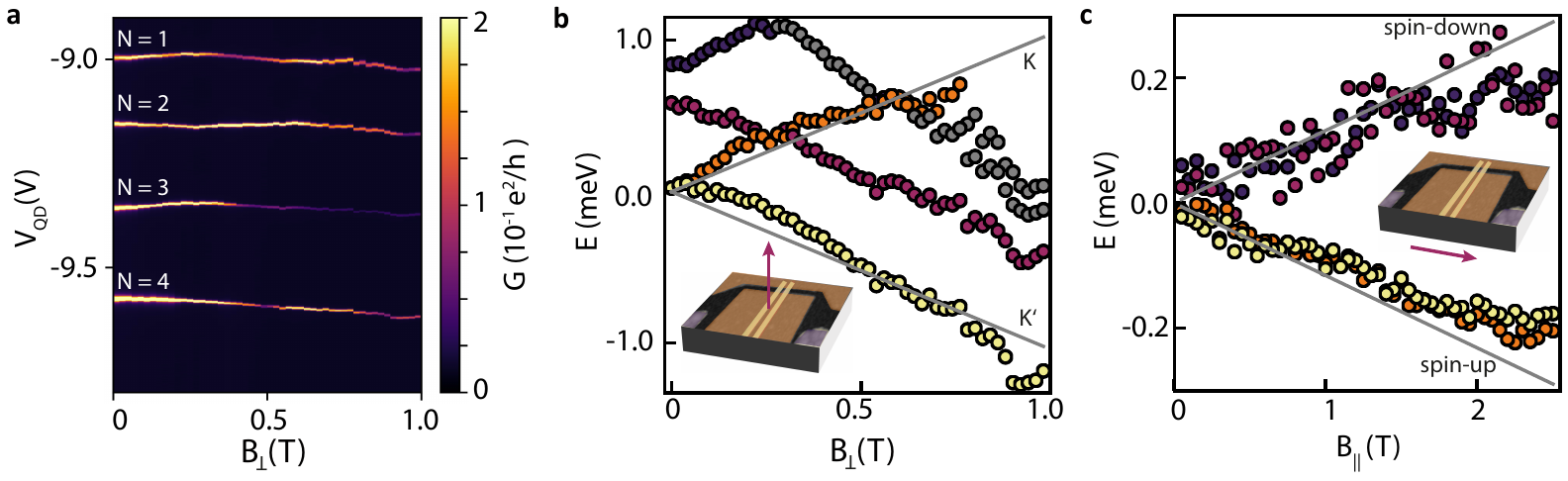}
	\caption{(a) Conductance map in a perpendicular magnetic field for the first Coulomb resonances of the QD in the hole regime. (b) Single particle energy level dispersion as function of the perpendicular magnetic field $B_\perp$ extracted from (a). The grey lines show a valley splitting with a valley g-factor $g_V=38$. (c) Parallel magnetic field dispersion for the four lowest single particle energy levels of the QD. Grey lines show the Zeeman splitting of free electrons.}
	\label{fig2}
\end{figure*}
The investigated sample schematically depicted in Fig.~\ref{fig1}(a) consists, from bottom to top, of a graphite back gate, a 33 nm thick hexagonal boron nitride (hBN) insulator, the bilayer graphene flake and another 35 nm thick hBN insulator, which are stacked using the dry transfer technique \cite{dean2010boron, uwanno2015fully}. Ohmic source- and drain-contacts are fabricated by edge-contacting \cite{wang2013one}. On top of this stack we deposit Cr/Au split gates (orange in Figs.~\ref{fig1}(a),(b)) defining a narrow channel. Separated by a 30 nm thick $\text{Al}_2\text{O}_3$ layer two Cr/Au finger gates are placed normal to the channel direction (yellow in Figs.~\ref{fig1}(a),(b), lithographic width 20 nm, gap 90 nm).

Back and top gates can be used (i) to open a band gap below the gates, and (ii) to tune the Fermi energy into the band gap, rendering these regions insulating. An n-type channel with a lithographic width of $100\,\text{nm}$ is formed between the split gates by applying a positive voltage to the graphite back gate ($V_{\text{BG}}=3.3\,\text{V}$) and a negative voltage to the split gates ($V_{\text{SG}}=-3.6\,\text{V}$). An in-plane source-drain bias voltage $V_{\text{SD}}$ is applied to the channel using the pair of ohmic contacts. 

A quantum dot is formed below the left finger gate in Fig.~\ref{fig1}(b) by accumulating holes with a negative finger gate voltage $V_{\text{QD}}$ \cite{eich2018spin, B}.  
Between n-type leads and the p-type dot the Fermi-energy traverses the band gap forming natural tunnel barriers for the quantum dot \cite{eich2018spin, B,doi:10.1021/acs.nanolett.8b01859,kurzmann2019charge}. All measurements were performed in a dilution refrigerator at an electronic temperature of 60 mK in a two-terminal DC setup with voltages $+V_{\text{SD}}/2$ applied to source, and $-V_{\text{SD}}/2$ to drain.

In Fig.~\ref{fig1}(c) we show differential conductance ($\partial I/\partial V_{\text{SD}}$) data of such a quantum dot measured in the few-hole regime. We label each diamond of suppressed conductance with the occupation number of the dot and extract an addition energy of 5 meV for the first hole and the lever arm of the finger gate $\alpha=0.029$. Excited states marked by arrows are observed running in parallel to the ground-state charging line. We use such spectroscopy measurements to extract the excited state energy spectra for one and two holes later in the paper.

Similar to Ref.~\cite{eich2018spin} we determine the valley ($g_V$) and spin ($g_S$) $g$-factor in the dot by measuring conductance resonances as a function of perpendicular and parallel magnetic field. In Fig.~\ref{fig2}(a) we show the first four conductance resonances in a perpendicular magnetic field with $V_{\text{SD}}<k_B T$. We extract the ground-state energy spectrum for varying occupation numbers of the dot in Fig.~\ref{fig2}(b), by subtracting charging energies (assumed to be magnetic field-independent) from the spacing of the resonances. The same procedure, applied to the data obtained in parallel magnetic field, results in the spectrum shown in Fig.~\ref{fig2}(c). In both Figs.~\ref{fig2}(b) and (c) changes of $V_{\text{QD}}$ were converted to energy using the lever arm $\alpha$. Comparing Figs.~\ref{fig2}(b) and (c) we find that the perpendicular magnetic field leads to a linear level splitting which is 19 times stronger than that in parallel field. In accordance with Ref.~\cite{eich2018spin} we interpret the parallel field splitting as the Zeeman effect and extract the expected g-factor for carbon materials $g_S = 2$. Similarly, we interpret the perpendicular field splitting as a valley splitting and extract $g_V = 38$ \cite{knothe2018minivalleys}.

The data in Fig.~\ref{fig2} suggest, that the filling sequence of the four degenerate single-particle states $\left|\left.K,s_z=\pm 1/2\right>\right.$, $\left|\left.K',s_z=\pm 1/2\right>\right.$ in the magnetic field range between 0 and 0.25 T is $\left|\left.K',s_z=+1/2\right>\right.$ (first hole, yellow in Fig.~\ref{fig2}(b) and (c)), $\left|\left.K,s_z=+1/2\right>\right.$ (second hole, orange), $\left|\left.K',s_z=-1/2\right>\right.$ (third hole, magenta), and $\left|\left.K,s_z=-1/2\right>\right.$ (fourth hole, dark blue),
thereby completing the occupation of the lowest energy shell. Similar shell occupation patterns are observed for higher occupation numbers (e.g., for 5-8 or 9-12 holes (not shown here)). This shell occupation pattern suggests, that the two-hole ground-state is a spin-triplet state ($s_z=1$), combined with a valley-singlet orbital wave function, whereas the three-hole ground-state has total spin $s_z=1/2$, and the four-hole ground-state has $s_z=0$.

 \begin{figure*}
	\includegraphics[scale=1]{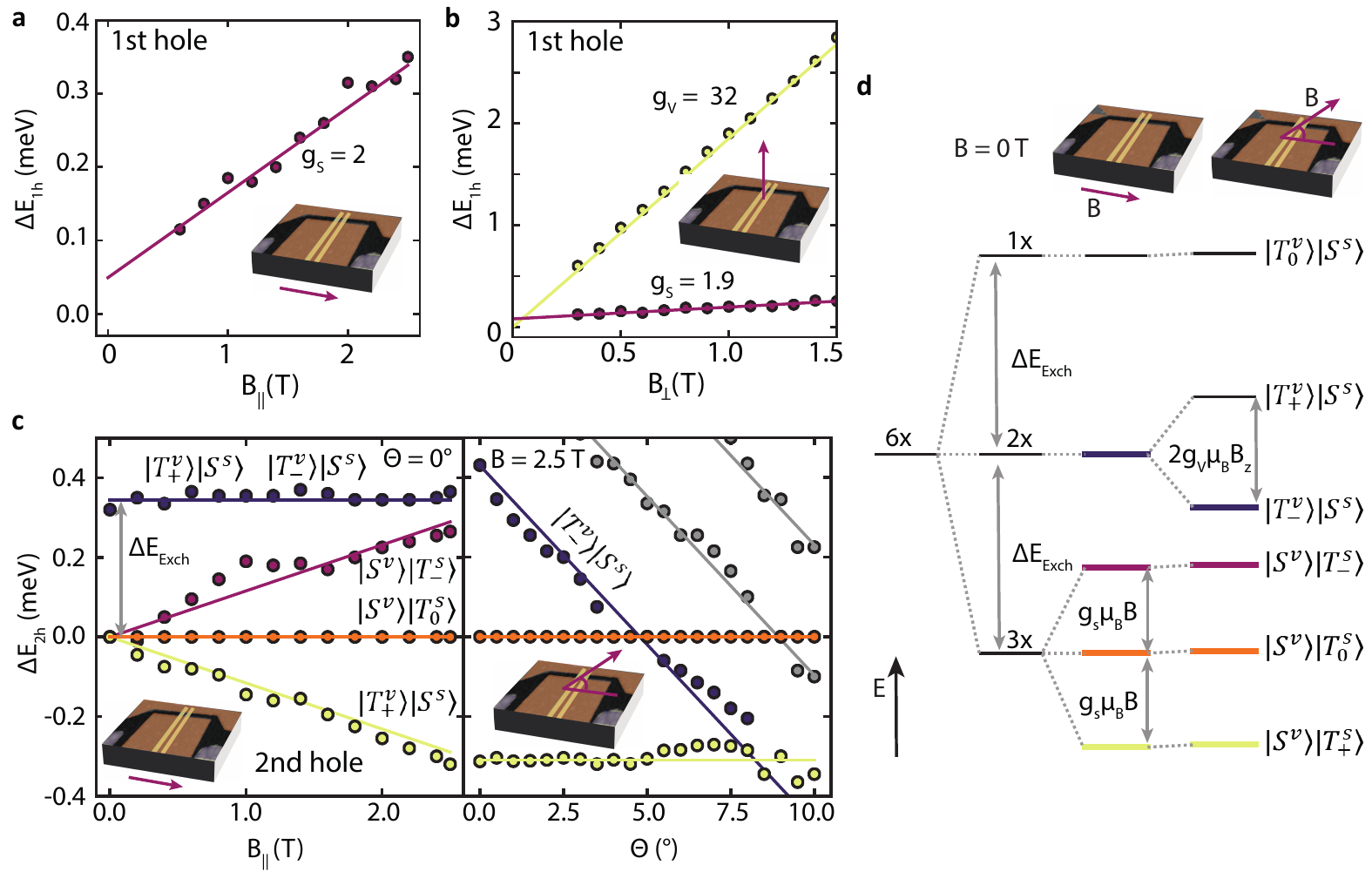}
	\caption{(a,b) Energy difference $\Delta E_{\text{1h}}$ between the ground-state and the excited states of the first hole occupying the dot in a parallel (a) and perpendicular (b) magnetic field.  (c) Left panel: Energy differences $\Delta E_{\text{2h}}$ between the states and the first excited state (orange) for the second hole occupying the dot in different magnetic fields applied parallel to the bilayer graphene flake. We identify two states that split with the magnetic field (yellow and magenta) and the energy of one excited state (dark blue) is independent of the magnetic field. Right panel: Excited state energies for different angles between the magnetic field axis and the bilayer graphen states. We observe one excited state that runs parallel to the ground-state (yellow) and three excited states (dark blue and grey), where the energy changes linearly with the angle $\Theta$. (d)  Energy levels for two holes occupying the QD. For vanishing exchange interaction and zero magnetic field all six two-hole states are energetically degenerate. A finite exchange interaction lowers the spin-triplet states by $\Delta E_{\text{exch}}$ in energy while the $\left| T_0^v \right\rangle\left| S^s \right\rangle$ state is raised in energy by $\Delta E_{\text{exch}}$. A finite magnetic field splits the spin triplet states by $g_S\mu_BB$, while a finite perpendicular component of the magnetic field splits mainly the valley triplet states by $g_V\mu_B B_\perp$.}
	\label{fig3}
\end{figure*}

Based on these findings, we take a closer look at the excited state spectrum of the one-hole system in parallel (Fig.~\ref{fig3}(a)) and perpendicular (Fig.~\ref{fig3}(b)) magnetic fields using spectroscopy measurements of the type shown in Fig.~\ref{fig1}(c). In Fig.~\ref{fig3}(a) and (b) we plot the energy difference between ground- and excited states and observe Zeeman (magenta) and valley (yellow) splitting of the lowest level. This observation confirms that the four single particle ground-states are degenerate at $B=0$ within experimental uncertainties. Extracted $g$-factors are indicated in the figure.

 
For two holes (see coloured arrows in Fig.~\ref{fig1}(c)) a set of four excited states is seen above the ground-state. In order to reveal the z-component of their spin, we show the $B_{||}$-evolution of these states  in Fig. \ref{fig3}(c, left panel). The ground-state is seen to Zeeman-split into three components (labeled $\left|\left.S^v\right>\right.\left|\left.T^s_-\right>\right.$,$\left|\left.S^v\right>\right.\left|\left.T^s_0\right>\right.$, $\left|\left.S^v\right>\right.\left|\left.T^s_+\right>\right.$), as expected for the proposed spin-triplet. The fermionic character of the total wave function dictates that these states have a valley-singlet wave function, denoted by $\left.\left| S^v\right.\right>$. The intensity of the $\left|\left.S^v\right>\right.\left|\left.T^s_-\right>\right.$ line, which is due to a tunneling transition from an $s_z=+1/2$ single-electron state to an $s_z= -1$ two-electron state is 8 times weaker than the intensity of the $\left|\left.S^v\right>\right.\left|\left.T^s_+\right>\right.$ line. In addition, a higher excited state is seen, which runs parallel to the $\left|\left.S^v\right>\right.\left|\left.T^s_0\right>\right.$ state. This state can either be a $\left|\left.S^v\right>\right.\left|\left.T^s_0\right>\right.$ spin state, or a spin-singlet state $\left|\left.S^s\right>\right.$.
 
 In order to disentangle these two options, we keep the magnetic field at 2.5 T, but tilt its orientation slightly out of plane by an angle $\theta$. The resulting level shifts are shown in Fig.~\ref{fig3}(c, right panel). As expected, the spin triplet states (yellow and orange) are insensitive to the tilt, because their spin-splitting depends on the total field only, while their valley-singlet character leads to neither orbital shifts nor splittings. The higher excited state (shown in dark blue), however, shows a strong angle dependence, identifying it as a valley triplet $\left|\left.T_-^v\right>\right.\left|\left.S^s\right>\right.$, and therefore a spin singlet state. We extract $g_V = 36$ for this state, in good agreement with the value found before for the ground-state transitions. Two additional excited states in Fig. \ref{fig3}(c, right panel, shown in grey) have the same valley triplet slope. We find the exchange energy $\Delta E_{\text{Exch}}=0.35\,\text{meV}$ from the energy separation between the $\left| T^v_- \right\rangle\left| S^s \right\rangle$ state and the $\left| S^v \right\rangle\left| T_0^s \right\rangle$ state at $B_z=0$.
 
 These experimental results are largely consistent with the two-hole level spectrum depicted in Fig.~\ref{fig3}(d). Six two-hole states can be constructed from pairs of the four degenerate single-particle states, degenerate in the absence of Coulomb interaction. Exchange interaction splits these states into a spin-triplet ground-state, three-fold degenerate at zero magnetic field, a spin-singlet state with a two-fold valley degeneracy at zero magnetic field, and a single spin-singlet valley-triplet state at the highest energy. Upon application of a parallel magnetic field, the triplet ground-state splits into its three spin-components, while all the other excited states remain unaffected. Adding a perpendicular magnetic field component (by tilting the field) splits the valley-triplet spin-singlet excited state, leading to a strong energy reduction of $\left|\left.T_-^v\right>\right.\left|\left.S^s\right>\right.$ with increasing tilt angle, in agreement with experiment. The two additional excited states observed in tilted magnetic fields (grey in Fig.~\ref{fig3}(c), right panel) are not captured within this level scheme.

The interpretation of our data is corroborated by a theoretical analysis of the Coulomb matrix elements of two-particle wave functions constructed from the single-particle model by Recher and co-workers \cite{PhysRevB.79.085407}. In the previous comparison \cite{eich2018spin} we identified the ground state in the absence of an applied magnetic field to have angular momentum number $m=1$. The resulting theory comparison gives an estimated value of $1\,\text{meV}$ for the exchange energy which is in good qualitative agreement with the measured value. It also shows that the off-diagonal matrix elements of the Coulomb interaction matrix are negligible for $m\neq 0$. The diagonal interaction matrix elements confirm the level spectrum shown in Fig.~\ref{fig3}(d).

In conclusion, we measured and identified the ground- and excited states of a few-hole quantum dot in bilayer graphene in a magnetic field. We identified a consistent level scheme for the lowest two-hole states of the dot. The ground-state at $B=0$ is a spin-triplet state, which can be viewed as Hund's rule for a valley-degenerate system and is in contrast to QDs in carbon nanotubes, where, as consequence of spin-orbit coupling, the two-particle ground-state is a spin singlet \cite{jarillo2004electron,huertas2006spin,kuemmeth2008coupling}. We, also extracted spin and valley $g$-factors and found them to be consistent between ground- and excited states. These experiments pave the way for measuring spin and valley relaxation and decoherence times of future graphene based qubits.

\section{Acknowledgments}
We thank Peter Märki, Erwin Studer, as well as the FIRST staff for their technical support. We also acknowledge financial support from the European Graphene Flagship, the Swiss National Science Foundation via NCCR Quantum Science and Technology, the EU Spin-Nano RTN network and ETH Zurich via the ETH fellowship program. FH is grateful for the financial support by the Swiss National Science Foundation through Division II (No. 163186 and 184739). Growth of hexagonal boron nitride crystals was supported by the Elemental Strategy Initiative conducted by the MEXT, Japan and the CREST (JPMJCR15F3), JST.

\end{document}